\documentclass[twocolumn,pra,showpacs,amsmath,superscriptaddress,floatfix]{revtex4-1}

\usepackage{booktabs} 
\usepackage{textcomp}
\usepackage{float}

\usepackage{epsfig,amssymb,amsmath,amsthm,amsfonts,amsbsy,mathrsfs,amscd}
\usepackage{graphicx}
\usepackage{color}

\usepackage[colorlinks,
            linkcolor= blue,
            anchorcolor= blue,
            citecolor= blue
            ]{hyperref}          
\usepackage{multirow}
\usepackage{booktabs}
\usepackage{lipsum}
\usepackage{tikz}
\usepackage{mathdots}
\usepackage{yhmath}
\usepackage{siunitx}
\usepackage{array}
\usepackage{gensymb}
\usepackage{tabularx}

\usepackage{makecell}
\begin{document}
\title{Comparisons among the Performances of Randomized-framed Benchmarking Protocols under T1, T2 and Coherent Error Models}
\author{Xudan Chai}\email{chaixd@baqis.ac.cn}\affiliation{Beijing Academy of Quantum Information Sciences, Beijing 100193, China}
\author{Yanwu Gu}
\affiliation{Beijing Academy of Quantum Information Sciences, Beijing 100193, China}
\author{Weifeng Zhuang}
\affiliation{Beijing Academy of Quantum Information Sciences, Beijing 100193, China}
\author{Peng Qian}
\affiliation{Beijing Academy of Quantum Information Sciences, Beijing 100193, China}
\author{Xiao Xiao}
\affiliation{Beijing Academy of Quantum Information Sciences, Beijing 100193, China}

\author{Dong E. Liu} \email{dongeliu@mail.tsinghua.edu.cn} \affiliation{State Key Laboratory of Low Dimensional Quantum Physics, Department of Physics, Tsinghua University, Beijing, 100084, China}\affiliation{Beijing Academy of Quantum Information Sciences, Beijing 100193, China}\affiliation{Frontier Science Center for Quantum Information, Beijing 100184, China}\affiliation{Hefei National Laboratory, Hefei 230088, China}

\begin{abstract}
While fundamental scientific researchers are eagerly anticipating the breakthroughs of quantum computing both in theory and technology, the current quantum computer, i.e. noisy intermediate-scale quantum (NISQ) computer encounters a bottleneck in how to deal with the noisy situation of the quantum machine. Since fully characterizing the quantum device has become technically impossible, and although an error mitigation technology has been adopted, it is still urgently required to construct more efficient and reliable benchmarking protocols through which one can assess the noise level of a quantum circuit that is designed for a quantum computing task. The existing methods that are mainly constructed based on a sequence of random circuits, such as randomized benchmarking (RB), have been commonly adopted as the conventional approach owning to its reasonable resource consumption and relatively acceptable reliability, compared with the average gate fidelity. To more deeply understand the performances of the above different randomized-framed benchmarking protocols, we design special random circuit sequences to test the performances of the three selected standard randomized-frame protocols under T1, T2, and coherent errors, which are regarded to be more practical for a superconductor quantum computer. The simulations indicate that MRB, DRB, and CRB sequentially overestimate the average error rate in the presence of T1 and T2 noise, compared with the conventional circuit's average error. Moreover, these methods exhibit almost the same level of sensitivity to the coherent error. Furthermore, the DRB loses its reliability when the strengths of T1 grow. More practically, the simulated conclusion is verified by running the designed tasks for three protocols on the Quafu quantum computation cloud platform. We find that MRB produces a more precise assessment of a quantum circuit conditioned on limited resources. However, the DRB provides a more stable estimation at a specific precision while a more resource-consuming.
\end{abstract}

\maketitle

\section{Introduction}
The emergence of increasingly powerful quantum technologies has brought millions of researchers in both experimental \cite{elben2023randomized,arute2019quantum,hangleiter2019sample}and theoretical\cite{chertkov2023characterizing} quantum physics into a new historical time node of the so-called third evolution of quantum through the developmental history of science and technology. This technological advancement has definitely triggered various unprecedented possibilities for exploring and exploiting the properties of complex quantum systems. As we know, either the exploration or the exploitation calls for more precise controllability\cite{torosov2022experimental} and manipulability\cite{altman2021quantum}. Simultaneously, this highly demanded technological development stimulates both the applications\cite{alexeev2021quantum} and the fundamental research \cite{kanno2023impact} within the quantum field, especially for the quantum computing we are discussing. 

We know the researchers belonging to this attractive field are now challenged by the complex, even unknowable noise within a quantum computer\cite{cheng2023noisy}, such as the superconductor quantum computer\cite{kim2023evidence}, i.e. a very popular platform for realizing real quantum computations. Due to its intrinsic complex operation mechanism, we are only capable of abstracting very limited information to the quantum circuit designed for performing a quantum computing task. Even worse, it is nearly practically impossible to fully and succinctly characterize 
a general large-scale error-corrected quantum computer, i.e. called Noisy intermediate-scale quantum (NISQ) computer\cite{bharti2022noisy} using classical data. Although error mitigation\cite{suzuki2022quantum,gu2023noise} as another selectable choice for developing the practical usability of a quantum computer has gained a lot of attention in this field, there are still urgent requirements both for experimentalists and theorists to theoretically create a more efficient and precise protocol\cite{gu2023benchmarking}, that can efficiently predict and precisely assess the quantum computer's performance with highly reasonable and practical reliability, especially for a large-scale quantum computer.

As far as we know, randomized benchmarking(RB) \cite{proctor2017randomized, gambetta2012characterization}together with various modifications of this method, such as interleaved RB \cite{magesan2012efficient}, direct RB \cite{proctor2019direct}, cycle RB \cite{erhard2019characterizing}, character RB \cite{zhang2023scalable}, mirror RB \cite{proctor2022scalable}, and even cross-entropy benchmarking(XEB) \cite{barak2020spoofing}, i.e., this method has been also framed into the structure of the general RB, has been preferred as the most practically reliable method recently. Although these protocols have been proven to be reliable and efficient based on numerical simulations and real experiments, a much clearer interpretation of their performances under more practical noises, such as T1, T2, \cite{baum2021experimental} and coherent errors \cite{duan1997preserving} of 1 and 2 qubit gate. 
There are a number of simulations and real experiments that have verified the reliability and efficiency of these protocols, however, it is still not very clear how they perform under more practical noises \cite{knill2005quantum}, such as T1, T2, and coherent errors of 1 and 2 qubit gate.

Although we have very limited information about the current superconductor quantum computer, we can still confirm that T1, T2, and coherent errors are the most significant noise resources that may unpredictably cause a catastrophic collapse \cite{harris2022benchmarking}for a quantum circuit system. 

Here, we select DRB, CRB, and MRB as three standard methods among the randomized-framed protocols \cite{helsen2022general} to test their performances under the two most concerned noises including T1, T2, and coherent errors for 1 and 2-qubit gate respectively within a whole quantum circuit system. Formally, these three different methods would probably contain not very much comparability due to the distinct practical purpose on which they are constructed. In order to contrast the performances of these methods, it is logically demanded that we carefully design a specific task that is operable for all of them and is also able to indicate the consistency or the difference among them. Therefore, we especially run a sequence of considerably designed random circuits \cite{harrow2009random} to test the performance of these three methods. The simulations show that in the presence of T1 and T2 noise, MRB, DRB, and CRB sequentially overestimate the average error rate of the circuit, compared with the circuit's average error rate calculated using the standard process tomography method. Moreover, the sensitivity \cite{proctor2022measuring}of these methods to the coherent error is nearly the same as the average gate fidelity calculated using the standard process tomography method. Furthermore, the DRB will fail especially with the strong T1 and T2 noise. 

Finally, in order to verify our simulated conclusion, we run the specifically designed task with these three methods on the Quafu quantum computation cloud platform\cite{group2023quafu,group2023quafu2,quafu}, which was released very recently. Except for the expected result, we also find that MRB produces a more precise assessment of a quantum circuit conditioned on limited resources. Conversely, aiming for a specific estimation precision, though more resource is consumed, the DRB provides a more stable estimation. In summary, the simulation and experiment demonstrated here contribute to the research in quantum benchmarking, therefore, we hope our work can be helpful in deeply understanding the characteristics of these protocols.

\section{Theory and Method}
Quantum computing has gained huge attention and even economic investments in the past decade years. There is a large growing number of experimental efforts and theoretic explorations that aim to demonstrate the reliability of scalable quantum computing. However, the current quantum computers suffer from unbearable errors, which are caused by complex situations, such as imprecise control, the natural decoherence of the qubits \cite{yoshihara2006decoherence}, and even cross-talk \cite{PhysRevLett.127.200502}. Obviously, a criterion \cite{knill2005quantum} is urgently needed to guarantee that the noisy scalable quantum computer is theoretically possible if the criteria are satisfied. One of the criteria is that the noise within physical gates is sufficiently low. Due to the complexity of the noise, the demonstration scalable quantum computer expects a particular type of noise that corresponds with the real situation as much as possible. The type of noise is often called an error model. So we must claim which type of error model is when we talk about the possibility. 

The threshold theorem \cite{knill2005quantum} indicates that criteria are currently a well-known rigorous guarantee that fault-tolerant quantum computing is possible if the threshold operating conditions are satisfied. The operating conditions are formed in two terms. One is that the noise must be promised from of noise, i.e. noise locality, and the other one is that the measurable errors must take place at a rate low enough within a quantum circuit. 
After roughly demonstrating the relationship between the error rate and the threshold theorem, we will explain these two terms in detail in the following two sub-parts.

\subsection{The Error rate of a quantum gate}
The error rate of a quantum is a measure that indicts the closeness of an operation of a noisy quantum gate to the ideal quantum operation. Usually, the above demonstration is clear enough to be understood. Think much deeper, we might find some insights to unfold the vague points within this concept. There are many metrics to determine the difference between operations, unfortunately, some have no operation meaning, namely not intuitively understandable, and some even have the wrong operation meaning. Here we choose distinguishability to be a pretty good operational foundation in our context. An error at a low enough level is required in fault-tolerant quantum computing, therefore, we are specifically interested in very small errors. However, according to quantum mechanics, it is very hard or even impossible to detect a single mistake with confidence. If all gates are perfect, the circuit will output the correct result, however, if we get a wrong result, this real gate will be distinguished from the ideal one. According to this, we can define the error rate to be the distinguishability of gates from the ideal. After all, this is just one way to define the error rate of an operation.

The statistics for an ideal process are governed by a probability distribution $p_{id}$, however, an error-prone produces a different distribution $p_{ac}$ that governs the actual statistics. The total variation distance \cite{verdu2014total}
\begin{equation}
    d_{TV}(p_{id},p_{id}) = \frac{1}{2}\Sigma_{x\in X }|p_{id}(x)-p_{id}(x)
\end{equation}
is a natural measure of the distance between two probability distributions over a set of outcomes $X $. Unfortunately, this measure is just theoretically meaningful while practically unmeasurable, especially for a large-scale qubit system. We all know that process infidelity drastically underestimates the distinguishability between unitary operations and the diamond norm distance \cite{shirokov2018energy}
\begin{equation}
    d_{\diamond}(\boldsymbol{G}_{ac},\boldsymbol{G}_{id}) : = \frac{1}{2} \left\|\boldsymbol{G}_{ac} - \boldsymbol{G}_{id}  \right\|
\end{equation}
between $\boldsymbol{G}_{ac}$ and $\boldsymbol{G}_{id}$ just provides an upper bound on distinguishability. 

One may hope that an estimate of the infidelity of the logical noise under a probabilistic Pauli channel generalizes directly general logical noise. Unfortunately, even quantifying the error becomes more complicated for more general noise. The ‘error rate’ of a noisy process $\varepsilon$ acting on a system is often experimentally quantified via the average gate infidelity to the identity, which can be efficiently estimated via randomized benchmarking\cite{knill2008randomized}. However, theoreticians often report rigorous bounds on the performance of a quantum circuit in terms of the diamond distance to the identity. The infidelity and diamond distance are related via the bounds\cite{15,16}. Pauli noise saturates the lower bound of and the effect of coherent noise is often assumed to be negligible so that experimental infidelities are often compared to diamond distance targets to determine whether fault tolerance is possible \cite{17}. However, even if coherent errors make a negligible contribution to the infidelity, they can dominate the diamond norm\cite {19}. Because of this uncertainty about how to quantify errors effectively, it is unclear what figure of merit recovery operations should optimize and how to quantify the logical error rate. \cite{PhysRevLett.121.190501} has proved that encoding a system in a stabilizer code and measuring error syndromes decoheres errors toward probabilistic Pauli errors.  Moreover, the error rate in a logical circuit is well quantified by the average gate fidelity at the logical level. 

\subsection{Metrics to the performance of quantum processor}
\subsubsection{Average gate fidelity}
The central task of quantum computation is to characterize the quality of quantum channels and quantum gates. The average gate fidelity of a quantum channel described by trace-preserving quantum operation $\varepsilon$ is defined by \cite{nielsen2002simple}:
\begin{equation}
  \overline{F}(\varepsilon)\equiv \int d\psi \left\langle \psi |\varepsilon (\psi)|\psi\right \rangle
\end{equation}
where the integral is over the uniform (Haar measure\cite{daele1997haar}) $d\psi$ on state space, normalized so $\int d\psi =1$. 
$\overline {F}(\varepsilon)$ can be further extended to a measure of how well $\varepsilon$ approximates a quantum gate, $U$,
\begin{equation}
  \overline{F}(\varepsilon, U)\equiv \int d\psi \left\langle \psi |U^{\dagger}\varepsilon (\psi)U|\psi\right \rangle
\end{equation}
where $\overline{F}(\varepsilon, U)=1$ if and only if $\varepsilon$ implements $U$ U perfectly, however, a lower value indicate that $\varepsilon$ is a noisy implementation of $U$. And $\overline{F}(\varepsilon, U)=\overline{F}(U^{\dagger}\circ\varepsilon)$,where $U^{\dagger}(\rho)\equiv U^{\dagger}\rho U$, and $\circ$ denotes composition of quantum gates.

Naturally, average gate fidelity is related to 
the gate error gate that we have defined. However, these two quantities are not so directly connected physically, or operationally. It means that the average gate infidelity $1-\overline{F}(\varepsilon, U)$ cannot be easily interpreted as an average error rate despite the common situation, i.e., because the measurement basis is not fixed in the integral so the infidelity is not an average error for a fixed measurement, yet neither is it averaged independently from the state. Instead, the lower and upper bounds to the error rate can be derived from fidelity, therefore, the mismatch of these two quantities or the substantial conditions with which they can directly connect with each other, are both in regimes of research interest. 

As authors in \cite{Sanders_2016} have clarified the possible relationship between these two quantities, information beyond fidelity is required to assess the relative importance of various noise processes that influence quantum devices. Pauli-distance defined in \cite{Sanders_2016} and unitarity defined in \cite{wallman2015estimating, kueng2016comparing} feature as two possible approaches to characterizing the influence of different noise sources. For example, if the Pauli-distance between an error channel and a Pauli channel \cite{flammia2020efficient}approaches zero, i.e. the error channel is close enough to a Pauli channel, $1-\overline{F}(\varepsilon, U)$ is directly connected with average gate error rate. Although no direct connection between them exists generally, average gate fidelity is still clearly of some worth: if the fidelity of quantum is precisely one, it is definitely certain that the gate will always perform exactly as expected.

\subsubsection{Entanglement fidelity}
Entanglement fidelity \cite{nielsen2002simple} as a quite simple and experimental useful formula is directly related to the average gate fidelity. To define this concept, we can assume $\varepsilon$ acts on one half of a maximally entangled state \cite{gisin1998bell}. Intuitively, if it acts on a qubit $Q$, and another qubit $R$
, with $RQ$ initially in the maximally entangled state labeled $\psi$. Then the entanglement fidelity can be defined to be the overlap between $\psi$ before and after the application of $\varepsilon$, 
$F_{e}(\varepsilon)\equiv\left\langle\psi(\boldsymbol{I}\bigotimes\varepsilon)(\psi)|\psi\right\rangle$, where $\boldsymbol{I}$ denotes the identity operation on system R. Thus the entanglement fidelity measures how well entanglement with other systems is preserved by the action of $\varepsilon$. Authors in \cite{nielsen2002simple} have put forward an elegant formula that connects $F_{e}(\varepsilon)$ to $\overline{F}(\varepsilon, U)$:
\begin{equation}
\overline{F}(\varepsilon, U) = \frac{dF_{e}(\varepsilon)+1}{d+1}
\end{equation}

In summary, either average gate fidelity or entanglement fidelity is useful for experimentally characterizing quantum gates and channels.

\subsection{The noises within a quantum circuit}
There are two main effects that cause the noises within a quantum circuit, one is the qubit interacting with the environment, and the other is the interaction between qubits.

An experimentalist of quantum computing always expects all the prepared qubits to be isolated, unfortunately, this expectation usually departs from the real case. The measurements or any non-unitary operations will affect the qubits to interact with the environment. The time of a qubit to keep isolated is coherent time, which yields two very important parameters in experiments. $T1$ as one of the parameters characterizes the time that a qubit relaxes from an excited state to a ground state, which is mostly caused by the interaction between the qubit system and the environment.

$T2$ as another important parameter determines the time that a qubit decay from a superposition state to either the excited state or the ground state.

Besides the above noises that result from the qubits interacting with the environment, quantum circuits always suffer from the disastrous crosstalk caused by the unwanted interaction with other qubits. A noise that is 
isolated is called a unitary error. Meanwhile, if not isolated, it might be the degradation of a quantum state, which is caused by imperfect operations or qubit leakage.


\subsubsection{Pauli twirling and randomized compiling}
The current superconductor quantum computers suffer from a noisy environment that is too complex to be precisely characterized. Technically, the noise within a quantum circuit can be converted into a specific quantum channel, i.e., twirling. Usually, we introduce a Pauli gate set to perform this particular twirling that is called Pauli twirling.  If a kind of noise can be represented as a matrix, and the effect caused by off-diagonal elements is ignorable, we can use Pauli twirling to convert this kind of noise into a Pauli noise channel.

Twirling \cite{ghosh2012surface, PhysRevA.72.052326}as a method is often used to approximate a general noise or even a complicatedly combined noise, with an asymmetric depolarization channel. We can use this procedure to study the average effect of arbitrarily general noise models by mapping them into more symmetric ones. Twirling over the Pauli group removes the off-diagonal terms, and even with the special condition satisfied, the asymmetric depolarization channel reduces to the symmetric depolarization channel. The above approximate reduction of any quantum channel to the asymmetric depolarization channel is usually referred to as the Pauli twirling approximation. As authors in \cite{PhysRevA.72.052326} proved that the standard forms of a completely positive map, the Pauli channel, and the depolarizing channel can be obtained by a random application of quantum operations before and after the actually completely positive map. These operations are chosen uniformly at random from a finite set of unitaries. Significantly, a depolarization twirling protocol does not introduce additional noise to the system, i.e. the Jamiolkowski fidelity remains the same, which can be regarded as some kind of distance measure that represents the noise level of the respective completely positive map and the standard form of it.

Coherent errors severely affect the performance of quantum computers in an unpredictable way. Hence, achieving a reliable quantum computation necessarily requires mitigating their severe impact. Worse while, the average error rates measured by randomized benchmarking and even similar protocols, highly lack sensitivity to the whole impact of coherent errors. Therefore, this makes the prediction of the global performance of quantum computers unreliable and makes it difficult to validate the accuracy of future large-scale quantum computations. Fortunately, a protocol called randomized compiling has been proposed to overcome these limitations by converting the coherent noise into stochastic noise. This protocol dramatically reduces unpredictable errors and allows us to accurately predict the performance of a quantum computer by measuring the error rates via cycle benchmarking. 

\begin{figure}
\centering
\includegraphics[width=8cm]{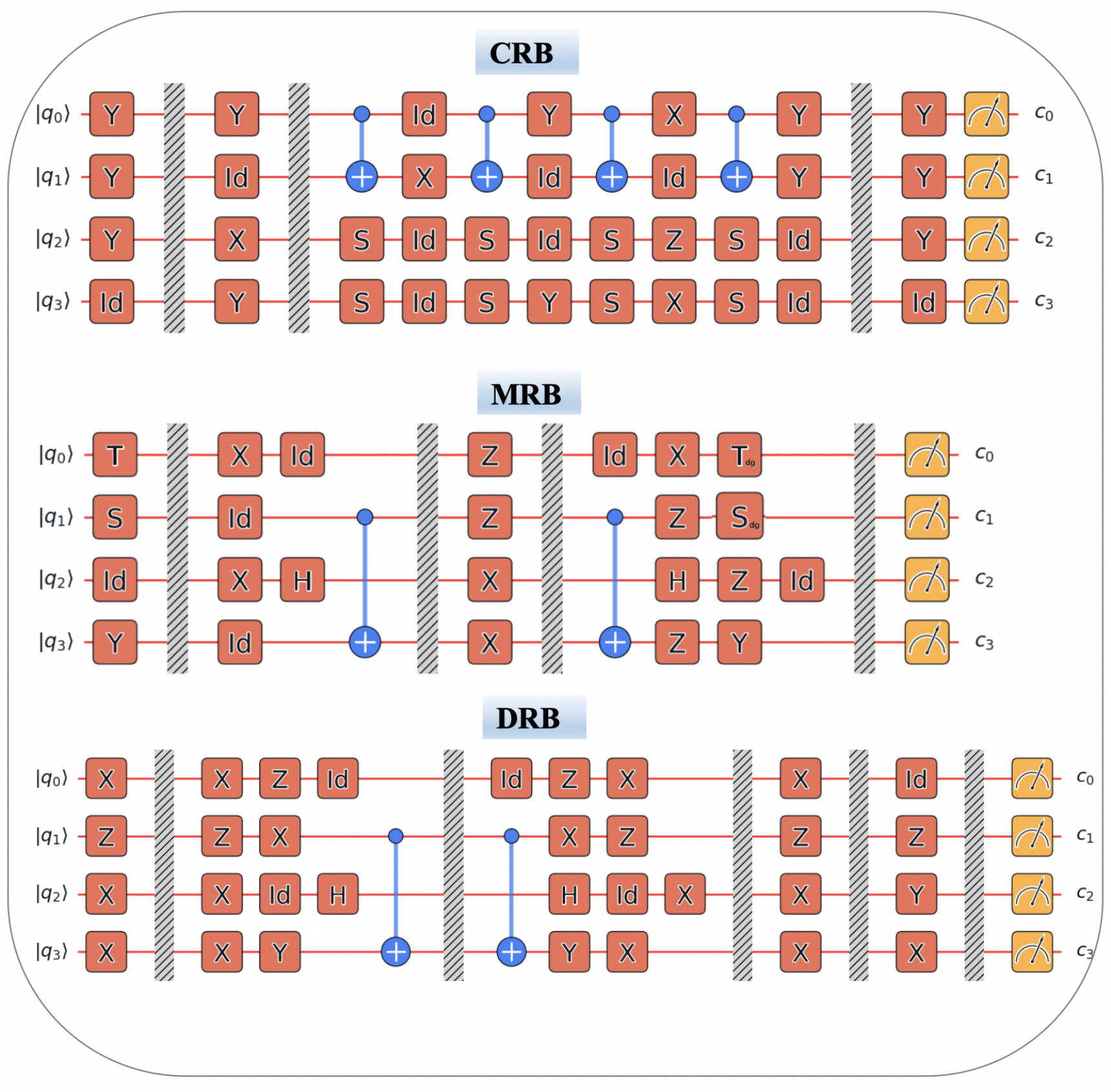}
\caption{Illustration of circuits required in Direct RB, Mirror RB, and Cycle RB.}
\label{circuit}
\end{figure}

Imagine a circuit that consists of single-qubit gate layers $C_k$ and two-qubit gate layers $G_k$, where $k$ is the index of the layers, we then insert and compile random single-qubit twirling gates\cite{hashim2020randomized}. $C_{k}\rightarrow T_{k}C_{k}T_{k-1}^c$, where $T_{k}$ is randomly sampled from a set of tensor products of single-qubit Paulis, and $T_{k-1}^c$ is chosen to undo the twirling gate, $T_{k}^c = G_{k} T_{k}^{\dagger}G_{k}^{\dagger} $. By this procedure \cite{hashim2020randomized}, the original single-gate layer will be logically equivalently replaced by the new layer without increasing depth instead.

There are following several major advantages of tailoring coherent errors into stochastic Pauli noise: (1) The off-diagonal terms in the error process caused by coherent errors can be completely suppressed(Pauli twirling assumed to be perfect); (2) Stochastic Pauli errors only grow linearly with circuit depth (in small error limit) because they occur with a finite probability in each gate cycle; (3) Stochastic Pauli errors have dramatically lower worst-case error rate than a coherent error at the same average rate. (4) The known fault-tolerant thresholds for stochastic noise are orders of magnitude higher than the threshold for generic local errors.

\section{Method}
\subsection{Protocols of Randomized-framed benchmarking}
As a practicably simple and efficient schema to estimate the noise level, conceptually, Randomized benchmarking (RB) proceeds as the following: (1). running random sequences of Clifford gates that are supposed to return the processor to its initial state(or a known randomized state.); (2). measuring the survival probability at the end of the circuit; (3). plotting the observed survival probabilities vs. sequence length and then fitting this to an exponential decay curve. The fitted decay rate of the survival probability that is up to a dimensionality constant allows  us to obtain the “RB number” -r, which is commonly as a metric for estimating the processor’s performance. Theoretically, RB can estimate a processor on any scale. However, it reaches its bottle because the Clifford group grows quickly with the number of qubits. Moreover, the procedure of compiling the Clifford gates into the native gates varies for different conditions of the hardware. The above is the operational burden and another one is the reliability of the RB number. It has been proven that the RB number precisely corresponds to the average gate infidelity when it is a Clifford group RB, however, it is not so clear for a general group structured RB. 

The drawbacks in terms of operability and reliability of RB yield a method called direct RB (DRB). The authors of DRB claim that an RB-number-like \textit{r} can be obtained by running many sequences of user-defined random circuits that consist of native gates directly instead of compiled ones. Moreover, the stabilizer state preparation that is used to encode the qubits brings another benefit, in which coherent errors can be converted into Pauli errors. Using this kind of technique, another method called cycle benchmarking (CRB) has been proposed and they claimed it to be reliable and efficient even for a non-Clifford gate set. Seems DRB and CRB have promoted efficiency and reliability obviously, unfortunately, the stabilizer preparation limits this method to apply to a larger scale qubit system. 

In order to solve the scalability of the previous protocols, the authors of MRB replace the 1-qubit Clifford state preparation rather than the multi-qubit stabilizer state. Besides this, they get the inversion of the user-defined random circuit by locally inverting each layer of the circuits, stead of the global inversion as the standard RB or DRM.
\newline
As benchmark protocols, DRB, CRB, and MRB shown in Fig.\ref{circuit} are structured based on random circuits, and the most practical thing is that we the users can define the random circuits according to special concerns. For instance, in order to determine the average error rate of a random circuit consisting of one and two-qubit gates, one can define a density  that determines the number of the two-qubit gates. Therefore, one can access the performance of this structural circuit by calculating the average error rates through sampling based on the probability distribution that has been defined. These three protocols can give us a parameter that indicates the performance of the specially defined circuits. This kind of parameter can be understood as fidelity but they are not the same things at all. 
\newline
\subsection{An instance defined to compare the performances of Randomized-framed Protocols} 
Since these protocols originally come from randomized benchmarking, the parameters exacted from the estimations would be operationally related to the average gate fidelity, however, these two parameters are not the same things essentially. Obviously, these protocols are operated very differently in terms of the structures of the circuits, and the formulas that are used to estimate the so-called error rates. Previous research has proved that these protocols are reliable for statistical noises and even coherent errors. However, it is not so clear for their performance under a more practical situation in superconductor qubit systems, for example, the T1 and T2 noise, and 1, 2-qubit gate coherent error. 
\newline
If we want to compare the performances of these three protocols, there will be several points that can not be ignored. Firstly, because there are a variety of parameters among these protocols, we have to control variables, such as (1) the structure that is defined by the density of two-qubit gates within the circuits; (2) the number of the repeated circuits that are required to guarantee the precision of estimations. Once the above two points have been satisfied, we can design an experiment to explore the performance of these protocols when the practical noises are considered.
\newline
In order to figure out the above concerns, we mainly choose the density of the 2-qubit gate as a significant changeable quantity to build a sequence of random circuits. Meanwhile, in order to make the comparison more practical,  we also consider the connectivity constraints in the generation of the designed random circuits because we know that the final structure of a superconductor quantum circuit is crucially determined by the topology structure of the hardware. Partially based on the definition in \cite{proctor2022measuring}, we define the density $\xi$ as $\xi = 2\alpha /wd$, where the $\alpha$ is the total number of the 2-qubit gates, and $w$ and $d$ are the 'width'(the number of qubits) and 'length' (the depth of a circuit) of a $w \times d$ lattice-shaped quantum circuit. With the above definitions, we demonstrate the procedure of generating random circuits as follows:

1) Assuming circuit with $\xi = 2\alpha /wd = 0.75$, the total number of the 2-qubit gates $\alpha$ equals $w\times d \times \xi /2 $;
\newline
2) For the circuit of the depth of $d$, the number of the 2-qubit gate (here we choose CNOT gate) for each layer is determined randomly with the connectivity constraints and the calculated $\alpha$ satisfied simultaneously; 
\newline
3) Once the 2-qubit gates are determined, the rest of the qubits that are idled will be filled with random 1-qubit Clifford gates from a set of $\left\{id, x, y, z, h, s, sdg, t, tdag\right\}$.
\newline
So far we have clarified the procedure to generate a sequence of random circuits and defined the crucial concepts associated. Now let's explain the reasonability with which we can compare the performances of these benchmarking protocols with the designed instance. According to the demonstrations in the second paragraph in the instance-design demonstration, it is clear that we have picked up the main factors of a quantum circuit at a gate level. Naturally, we choose the control variable method to those factors to make the results of comparison among those protocols more logically acceptable for the sake of rigor.
\begin{figure}
\includegraphics[width=8cm]{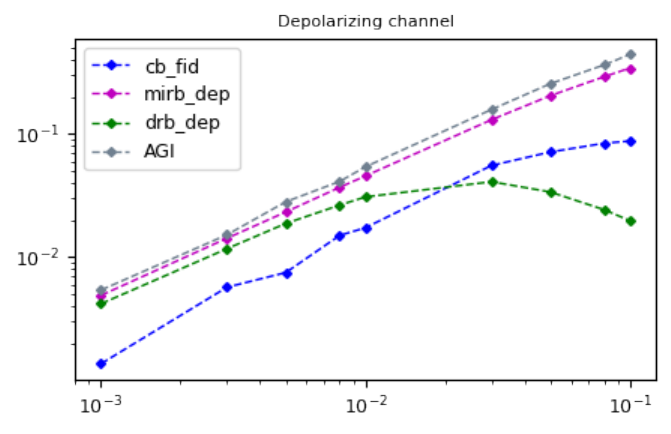}
\caption{The average gate error rate estimated by using MRB, DEB, CRB and process infidelity(i.e. equals to the average gate infidelity in the case of depolarization channel based on the \cite{Nielsen_2002}.)}
\label{dep}
\end{figure}
\section{Results of the Simulation and Experiment}
\vspace{1cm}
\subsection{Experimental results and discussion}
\vspace{1cm}
Even though we have clarified the reasonability of comparing MRB, DRB, and CRB, it is still necessary to test it with a very trial situation, i.e. depolarization channel, with which it is much clearer for the numerical calculation and the physical interpretation and as well as the possible consistency between them and the operational meaning. 
\begin{figure*}
\centering
\includegraphics[width=16cm]{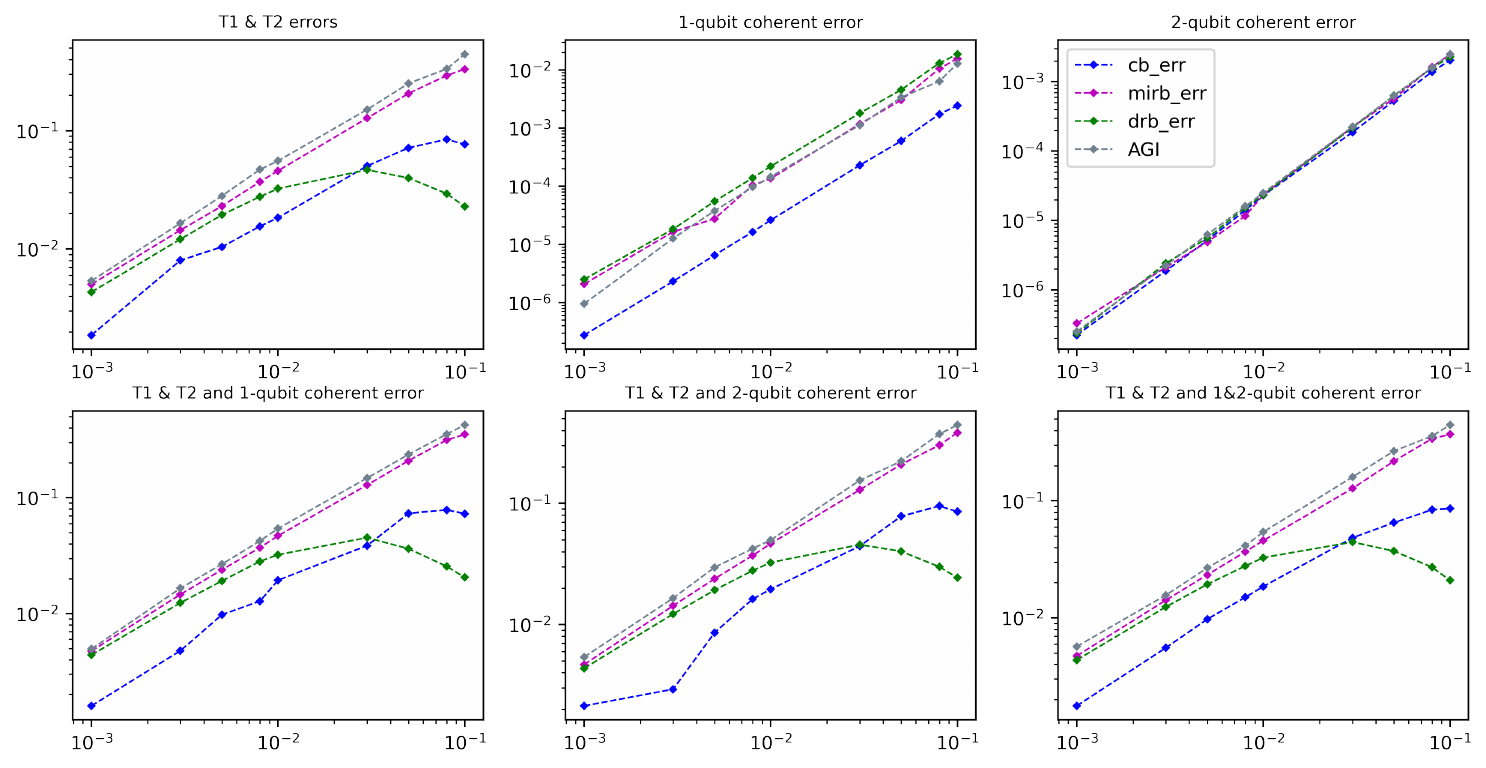}
\caption{The average gate error rates estimated by MRB, DRB, CRB, and the process tomography for a reference, under the T1, T2, and 1 and 2-qubit coherent errors.}
\label{rbs}
\end{figure*}
It might be sufficient for an experimentalist to perform a quantum computer benchmarking with only clear and efficient instructions. However, it might lack much further to figure out the deep operating mechanism of the procedure. For example, why and how can this protocol convince us? Or is there any possibility of generalizing this protocol for a more practical or general situation etc..... Thanks to the elegant results in \cite{Nielsen_2002}, we could have understood this family of protocols at a deeper theoretical level. It has been very clear for the interpretation of the RB when the case is depolarization with the noise assumptions satisfied, for which the average gate fidelity (infidelity) estimated through the original definition exactly equals the averaged error rate estimated through the whole procedure according to RB. Therefore, before we investigate their performances in different situations, we choose this clearly interpreted situation to test the intuitive concern of why we could perform this research by building this instance. According to the previous study, we can ensure that the validity of our research would be verified if the results obtained through these three protocols are consistent with those calculated through the 
rigorous process tomography. As expected, we get this consistency in our simulation, which is shown in Fig. \ref{dep}.
\begin{figure*}[t]
\includegraphics[width=17cm]{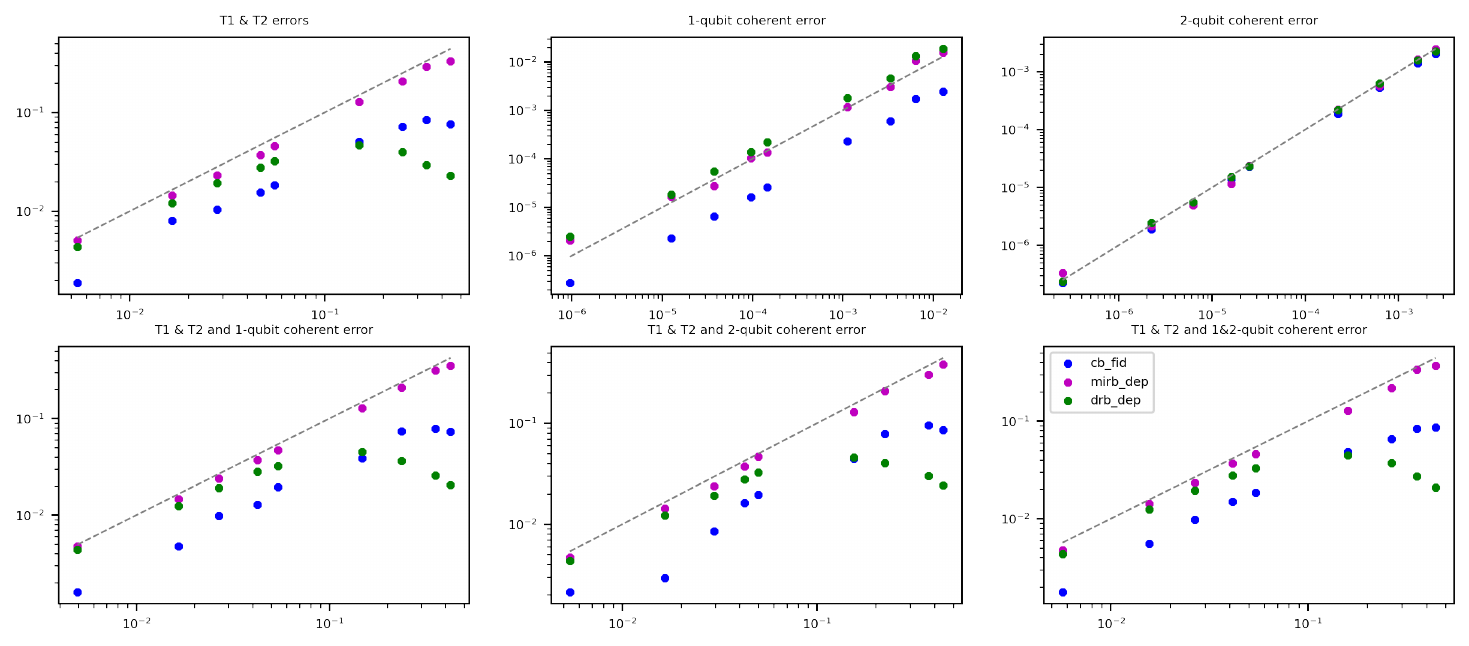}
\caption{A more intuitive display of the results shown in Fig. \ref{rbs}. The colorful dots represent the estimations of MRB, DRB, and CRB, and the gray dot line as a reference is obtained by process tomograph.}
\label{rbest}
\end{figure*}

As shown in Fig. \ref{rbs}, these three behave quite similarly when the noise strength is kept lower than $10E^{-2}$ (note that we compare only at the level of orders of magnitude ). However, the DRB will underestimate the noise level when the noise within a quantum circuit becomes worse and worse. Finding this, we will go further for more practical noisy situations, such as the T1, T2, and 1 and 2-qubit gate errors. Fig. \ref{rbs} shows the results under this situation. The upper panels are the case of only one kind of noise while the lower panels are the combinations of those noises. The first panel of T1 and T2 errors only shows that MRB, DRB, and CRB estimate similar error rates compared to the process-tomography estimations. The difference is that the DRB gives a misleading estimation when the strength of the T1 \& T2 errors is nearly larger than $10E^{-2}$, or we would rather say that it goes far from the real situations. 

\begin{figure}
\includegraphics[width=9cm]{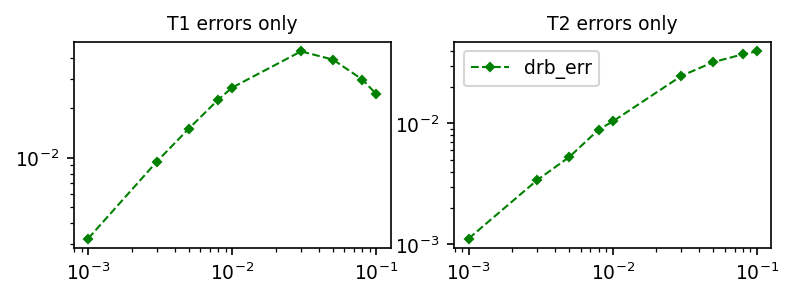}
\caption{The performances of DRB under the T1 and T2 errors. It is obvious that the T1 dominates the final behavior of the DRB.}
\label{drbcheck}
\end{figure}

The next two panels of Fig. \ref{rbs} show the results of the two kinds of coherent errors. If we regard the reaction to the variation in the strength of one kind of noise, as the sensitivity of a benchmarking protocol to that kind of noise, we can conclude that these three protocols exhibit nearly the same as the process tomography. It is worth mentioning that no differences we can find between the DRB and the other protocols. Having these results, we can go on with several possible combinations of these noises. The lower panels tell us one important information the performance of DRB is almost dominated by the T1 and T2 errors rather than coherent errors, no matter if it is a 1 or 2-qubit coherent error. Obviously, the special performance of the DRB requires a further understandable interpretation. Based on the procedure of the DRB, it requires preparing a stabilized state at the first step. 
\begin{figure}
\includegraphics[width=8cm]{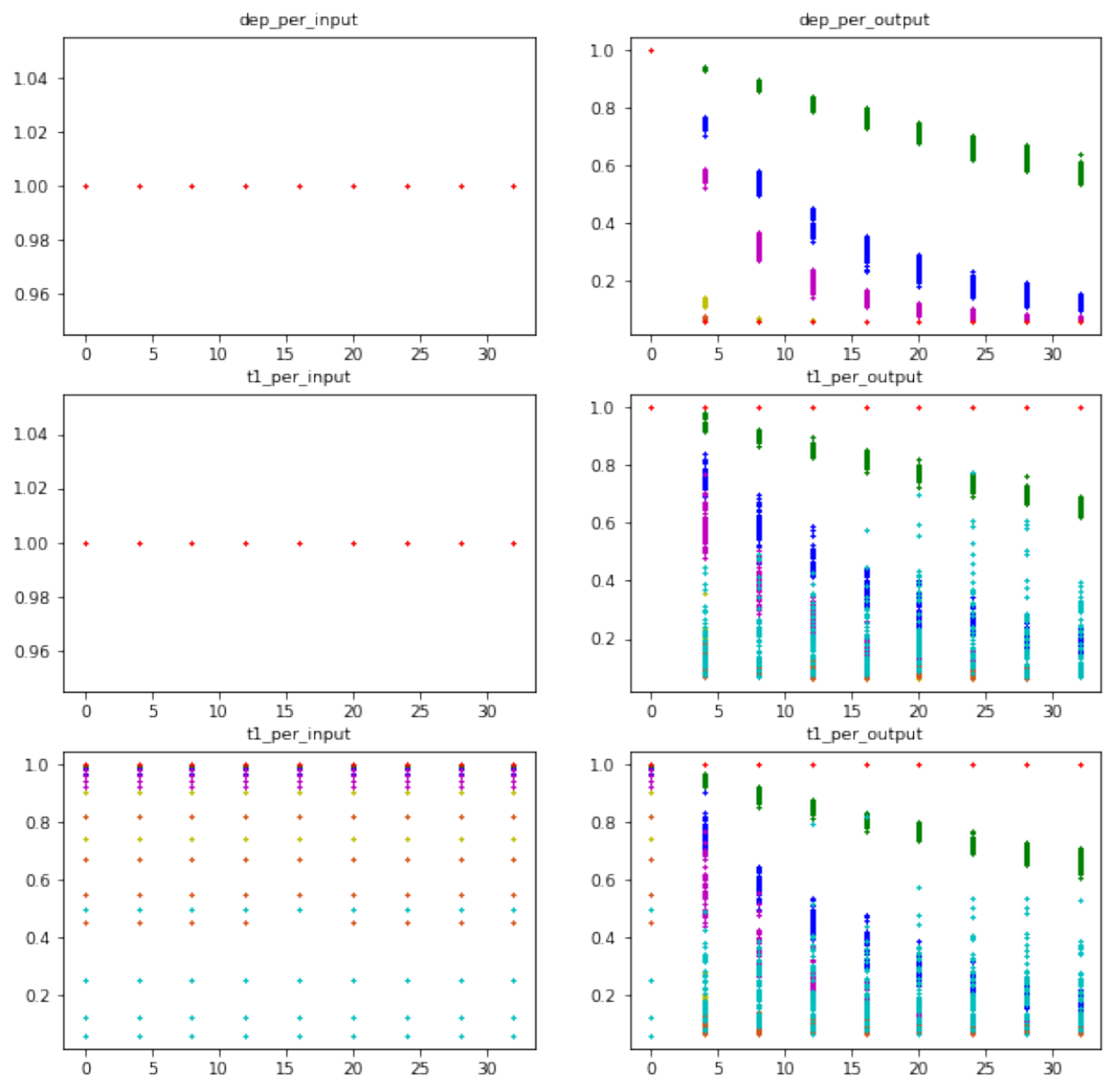}
\caption{The purity variation of the initial prepared states at the first stage and the output state at the final stage in a circuit benchmarking with the DRB method.}
\label{drbpurity}
\end{figure}
Correspondingly, the last step conversely performs the inverse of the first state preparation. Unfortunately, even though the circuit of this protocol has suffered a devastating collapse at the first state preparation, almost little information we can abstract only according to the measurement obtained under the computational basis. This explanation can be more evidence-based as shown in Fig. \ref{drbpurity}, which simulates the above guesses. In order to be more precise, we introduce the purity to test the suspect. Firstly, as shown in Fig. \ref{drbcheck}, we find the departure of DRB mainly resulted from the T1 errors instead of the T2 errors. Obviously, we know that the T1 errors that result from the interaction with the environment would scramble the expected initial state into a mixed one. And then combined with Fig. \ref{drbpurity}, it is not difficult to imagine that the final estimation would exact no more extra information about the quantum circuit than the first state preparation stage. Therefore, we can say that only a very precise initial stabilized state preparation would give us an accurate estimation of the noise level of the circuit for the DRB protocol. However, we know that this requirement is usually too harsh for the sake of practicality.
Furthermore, we replot the results into Fig \ref{rbest} in order to make it more directly compared among these protocols. It is clearly shown in Fig \ref{rbest} that all three kinds of randomized-framed protocols understand the noise level compared to the process tomography.
\begin{table}
\setlength{\tabcolsep}{0.05mm}
\begin{tabular}{|c|c|c|c|c|}
\hline
Chip & $T1$ & $T2$ & \text{$Fidelity_{cnot}$} & Connectivity \\
\hline
\multirow {3}{*} & Avg: 34.208 & Avg: 17.485 & Avg: 0.946 & Square \\
ScQ-P136 & min: 15.17 & min: 0.95 & min: 0.83 & Matrix \\
~& max: 59.1 & max: 53.41 & max: 0.996 & Neighbors \\
\hline
\end{tabular}
\caption{The details of calibration information of ScQ-P136.} 
\label{table1}
\end{table}
\subsection{Experimental results and discussion}
So far one may ask, since the displayed results are nearly the same except the extreme situations, can it more clear about the resource consumption of these protocols, as well as the possible precision of the estimations with given only a very limited resource, such as the total quantum gates consuming. With the above concerns, we test our instance on a real quantum computing platform, which is called the QuaFu quantum cloud platform. We choose the \textit{ScQ-P136} chip \ref{table1}to run the instance that has been numerical simulated on in \textit{qiskit}.


Finally, we implement the experiment on the Quafu quantum cloud platform. As shown in \cite{quafu}, we find that these three protocols estimate the error rate when the number of qubits is small. However, when it multiplies,  the error bars increase. 
\begin{figure}
\includegraphics[width=7cm]{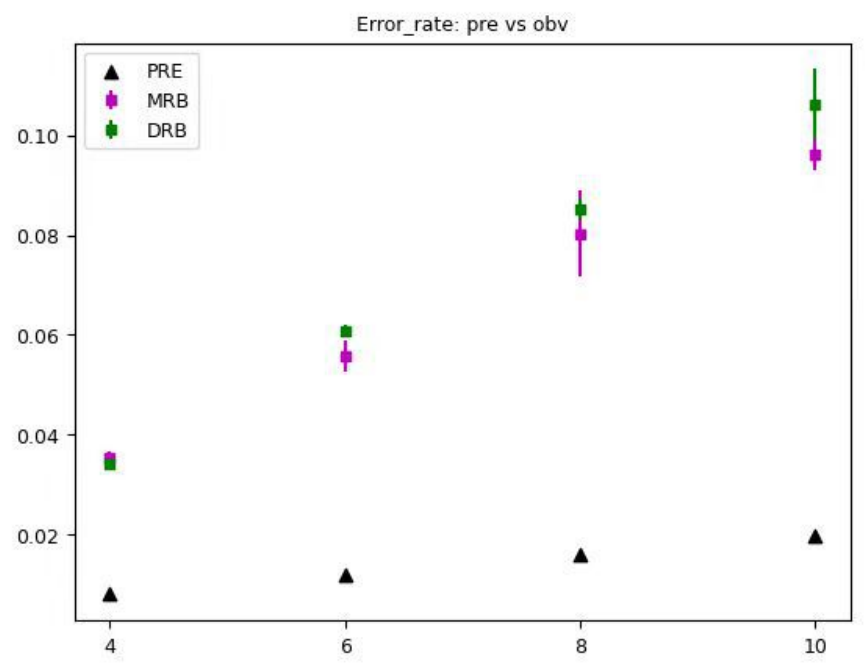}
\caption{The experiments of quantum circuits benchmarking on the Quafu platform using MRB and CRB, and the black triangles are the most optimist estimation of the error rate.  }
\label{qf}
\end{figure}

\section{conclusion}
In conclusion, the field of quantum computing holds tremendous potential, fueling the anticipation of groundbreaking advancements in both theory and technology. However, the current quantum computing landscape, characterized by noisy intermediate-scale quantum (NISQ) computers, confronts a formidable obstacle in managing quantum machine noise. The complete characterization of these quantum devices remains a technical impossibility, even with the adoption of error mitigation technologies.

Addressing this challenge, there is a pressing necessity to develop more efficient and dependable benchmarking protocols capable of evaluating noise levels in quantum circuits designed for specific computational tasks. Established techniques, predominantly rooted in random circuit sequences such as randomized benchmarking (RB), have gained widespread acceptance due to their judicious resource utilization and satisfactory reliability compared to average gate fidelity.

To attain deeper insights into the performance of these benchmarking protocols, specialized random circuit sequences were meticulously crafted to assess three standard randomized-frame protocols under the influence of T1, T2, and coherent errors—circumstances more pertinent to superconducting quantum computers. Simulations disclosed that MRB, DRB, and CRB consistently overestimated the average error rate in the presence of T1 and T2 noise when juxtaposed with conventional circuit average errors. Furthermore, these methods displayed akin sensitivities to coherent errors. Remarkably, the reliability of DRB waned as T1 strengths increased.

To substantiate the practicality of these findings, the simulated conclusions were authenticated by implementing designed tasks for the three protocols on the Quafu quantum computation cloud platform. The outcomes underscored that MRB furnished a more precise assessment of a quantum circuit within resource constraints, while DRB proffered a more steadfast estimation at specific precision levels, notwithstanding its greater resource utilization.

These research findings elucidate the intricacies of noise mitigation in quantum computing and appraise the efficacy of benchmarking protocols under diverse error scenarios. They enrich our comprehension of the constraints and avenues for enhancement in NISQ quantum computers, thus paving the way for more dependable and efficient quantum computing technologies in the future.

\section*{Acknowledgments} We would like to thank Dr. Bo Gao at the Beijing
Institute of Technology for her careful revision of the manuscript. This work was supported by
the Beijing Natural Science Foundation (No. Z220002). 
\bibliography{manuscript}
\end{document}